\documentclass[twocolumn,showpacs,preprintnumbers,amsmath,amssymb]{revtex4}
\usepackage{tabularx,graphicx}\begin{document}
\newcommand{\beq}{\begin{equation}}
\newcommand{\eeq}{\end{equation}}
\newcommand{\beqn}{\begin{eqnarray}}
\newcommand{\eeqn}{\end{eqnarray}}
\newcommand{\bmath}{\begin{subequations}}
\newcommand{\emath}{\end{subequations}}
\title{Double-valuedness of the electron wave function and rotational zero-point motion of electrons in rings }
\author{J. E. Hirsch }
\address{Department of Physics, University of California, San Diego,
La Jolla, CA 92093-0319}

\begin{abstract} 
I  propose that the phase of an electron's wave function changes by $\pi$ when the electron goes around a loop  maintaining phase coherence. 
Equivalently, that the minimum orbital angular momentum of an electron in a ring  is $\hbar/2$ rather than zero as generally assumed, hence that the electron
in a ring has azimuthal zero point motion.
This proposal provides  a physical explanation for the origin of electronic  `quantum pressure', 
it  implies that a spin current exists in the ground state of aromatic ring molecules, and it suggests an explanation for the ubiquitousness
of persistent currents observed in  mesoscopic rings.

    \end{abstract}
\maketitle

\section{introduction}
   \begin{figure}
 \resizebox{8.5cm}{!}{\includegraphics[width=8cm]{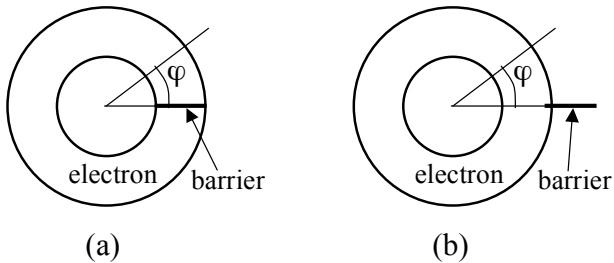}}
 \caption  { Electron in a ring.  In (a), an infinite potential barrier exists at azimuthal angle $\varphi=0$, forcing the electron wavefunction to be zero there. In (b) the barrier is removed.
}
 \label{figure2}
 \end{figure}

Quantum mechanics teaches us that when we confine a particle to a small spatial region it acquires ``zero point motion'', the more so the smaller the region.
Consider an electron in the ring shown in Fig. 1(a), with an infinite potential barrier at azimuthal angle $\varphi=0$. According to 
Schr\"odinger's equation, its
ground state wave function is given by
\beq
\psi(\varphi) \propto sin(\varphi/2)=\frac{e^{i\varphi/2}-e^{-i\varphi/2}}{2i}
\eeq
as far as its azimuthal dependence is concerned, to satisfy the boundary condition
\beq
\psi(\varphi=0)=\psi(\varphi=2\pi)=0
\eeq
where the potential is infinite. Its azimuthal energy (kinetic energy from azimuthal degree of freedom) is
\beq
E_\varphi=\frac{\hbar^2\pi^2}{2m_e P^2}
\eeq
with $P=2\pi r$, $r$ the (average)  radius of the ring and $m_e$ the electron mass. The zero point energy Eq. (3) arises from the confinement of the electron in the perimeter length 
$P$, or equivalently from the
confinement of the azimuthal angle $\varphi$ in the finite region
\beq
0\leq \varphi \leq 2\pi
\eeq
as ordained by the uncertainty relations $\Delta l \Delta p_l \sim \hbar$ ($l$ being the arc length, $p_l$ the associated linear momentum), or $\Delta \varphi \Delta L_z \sim \hbar$, with $L_z$ the angular momentum perpendicular to the
plane of the ring.

Ordinarily an electron confined to a finite box exerts ``quantum pressure'' on the walls, and moving a wall outward lowers the quantum zero point energy. 
In Fig. 1(a) however,  equal ``quantum pressure'' is exerted on both sides of the potential barrier so it averages to zero. This suggests that 
{\it the energy will not be lowered if the barrier is removed}. 
 However, conventional quantum mechanics predicts that when the barrier is completely removed 
 (Fig. 1(b)) the ground state wavefunction is
\beq
\psi(\varphi)=constant
\eeq
with azimuthal energy 
\beq
E_\varphi=0, 
\eeq
hence that  the energy is lowered by
$\Delta E=-\hbar^2\pi^2/(2m_eP^2)$  from the case where the barrier is completely in. One may resonably ask, how is the electron able to exert a radial  `force' on the barrier to
 lower its energy from Eq. (3) to Eq. (6) in the geometry of figure 1?
 
 Note also the following peculiarity of the wavefunction Eq. (1). Its average angular momentum in the $z$ direction 
 $<L_z>=0$, and its 
 uncertainty is
 \beq
 \Delta L_z=\sqrt{<L_z^2>-<L_z>^2}=\frac{\hbar}{2}  .
 \eeq
 This suggests that a measurement of the $L_z$ angular momentum of the electron will yield half of the time $\hbar/2$, the
 other half of the time $-\hbar/2$. Usually in quantum mechanics when a particle is in a 
 state which is a coherent superposition of states with well-defined quantum numbers, there is a way to `collapse' the
 wavefunction (e.g. by doing a measurement) into one of the states with one value of the quantum number. 
 However within conventional quantum mechanics there is no  way to do that  with the electron in the wavefunction Eq. (1).
 Instead, when the barrier is removed, the wavefunction `collapses' to Eq. (5) with  angular momentum
 $L_z=0$ and no uncertainty. What happened to the fluctuating angular momentum?
 
 Another peculiarity of the wavefunction Eq. (1) is that it is double-valued, namely $\Psi(\varphi+2\pi)=-\Psi(\varphi)$.
 Within conventional  quantum mechanics this double-valuedness is interpreted as having no physical significance
 because the electron cannot go `across' the barrier. But could it acquire significance when the barrier is removed?
 
 Let us analyze what happens to the electron in Fig. 1(a) as the barrier is gradually removed within conventional
 quantum mechanics. Consider a variational wavefunction of the form
 \beq
 \Psi(r,\varphi)=f(r)sin(\varphi/2)
 \eeq
 with $r$ the radial coordinate. Of course Eq. (8) cannot be the true wavefunction because it has discontinuous derivative at points
 where the potential is not infinite (radial positions where the barrier is absent and $\varphi=0$). Nevertheless, it is well known that variational wavefunctions with 
 piecewise continuous first derivatives are allowed within the variational principle and yield valid upper bounds to the ground state energy\cite{parr}. Thus, the energy Eq. (3) is an upper bound for  the azimuthal contribution to the  ground state energy for any
 position of the barrier in the ring.
 
 We can find a better upper bound for the case when the barrier is almost completely out. Let
 $r_1$, $r_2$ be the inner and outer radii of the ring, and $w=r_2-r_1$ the ring width. Assume the
 barrier occupies the region $r_2-\Delta w\le r\le r_2$ for $\varphi=0$. The true wavefunction goes to zero at
 $r=r_1$ and $r=r_2$ for any $\varphi$. Take as variational wavefunction one that is confined to the radial region
 not occupied by the barrier ($r_1<r<r_2-\Delta w$) and constant as function of $\varphi$, and identically zero in the
 region $r_2-\Delta w \le r \le r_2$ for all $\varphi$. Its energy is
 \beq
 E_1=\frac{\hbar^2}{2m_e}(\frac{\pi^2}{(w-\Delta w)^2})
 \eeq
 while the energy when the barrier is completely in, which is also an upper bound for any position of the barrier, is
 \beq
  E_{in}=\frac{\hbar^2}{2m_e}(\frac{\pi^2}{w^2}+\frac{\pi^2}{P^2})   .
 \eeq
 Thus, Eq. (9) is a better upper bound than Eq. (10) when $E_1<E_{in}$, i.e. 
 \beq
 \frac{\Delta w}{w}\le \frac{1}{2} (\frac{w}{P})^2   .
 \eeq
 According to this variational wavefunction, the electron exerts a (nearly) constant outward force
 $-\partial E_1/\partial \Delta w$ on the barrier when the barrier is almost completely out.
 
 To  gain further insight, we solve the  Schr\"odinger  equation numerically on a discrete lattice with $N_r$ points in the radial
 direction and $N_\varphi$ points in the azimuthal direction. The boundary conditions are $\Psi(r_1,\varphi)=\Psi(r_2,\varphi)=0$
 and $\Psi(r,0)=0$ for $r_2-\Delta w\le r\le r_2$, where the barrier is. The lattice spacings are
 \bmath
 \beq
 a_r=\frac{w}{N_r+1}
 \eeq
 \beq
 a_\varphi=\frac{P}{N_\varphi+1}
 \eeq
 \emath
 with $P=\pi(r_1+r_2)$ the average perimeter of the ring.
 The nearest neighbor hopping amplitudes in the $r$ and $\varphi$ directions are
 \bmath
 \beq
 t_r=\frac{\hbar^2}{2m_e}\frac{1}{a_r^2}=\frac{\hbar^2}{2m_e}\frac{(N_r+1)^2}{w^2}
 \eeq
  \beq
 t_\varphi=\frac{\hbar^2}{2m_e}\frac{1}{a_\varphi^2}=\frac{\hbar^2}{2m_e}\frac{(N_\varphi+1)^2}{P^2}
 \eeq
 \emath 
 When the barrier is fully in, the ground state wavefunction is
 \beq
 \Psi(i_r,i_\varphi)=f(i_r)g(i_\varphi)
 \eeq
 with
 \bmath
 \beq
 f(i_r)=\sqrt {\frac{2}{N_r+1}}sin(\frac{\pi  i_r}{N_r+1})
 \eeq
  \beq
 g(i_\varphi)=\sqrt {\frac{2}{N_\varphi+1}}sin(\frac{\pi  i_\varphi}{N_\varphi+1})
 \eeq
 \emath
 and $i_r$, $i_\varphi$ labeling the discrete lattice points in the $r$ and $\varphi$ directions. When the barrier is
 completely out, the ground state wavefunction is Eq. (14) with $f(i_r)$ given by Eq. (15a) and 
 $g(i_\varphi)=1/\sqrt{N_\varphi}$. The ground state energies are given by
 \bmath
 \beq
 E_{in}=\frac{\hbar^2}{2m_e}(\frac{\pi^2}{w^2}+\frac{\pi^2}{P^2}) 
 \eeq
  \beq
 E_{out}=\frac{\hbar^2}{2m_e}(\frac{\pi^2}{w^2}) 
 \eeq
 \emath
  when the barrier is fully in and out respectively.
  
     \begin{figure}
 \resizebox{8.5cm}{!}{\includegraphics[width=8cm]{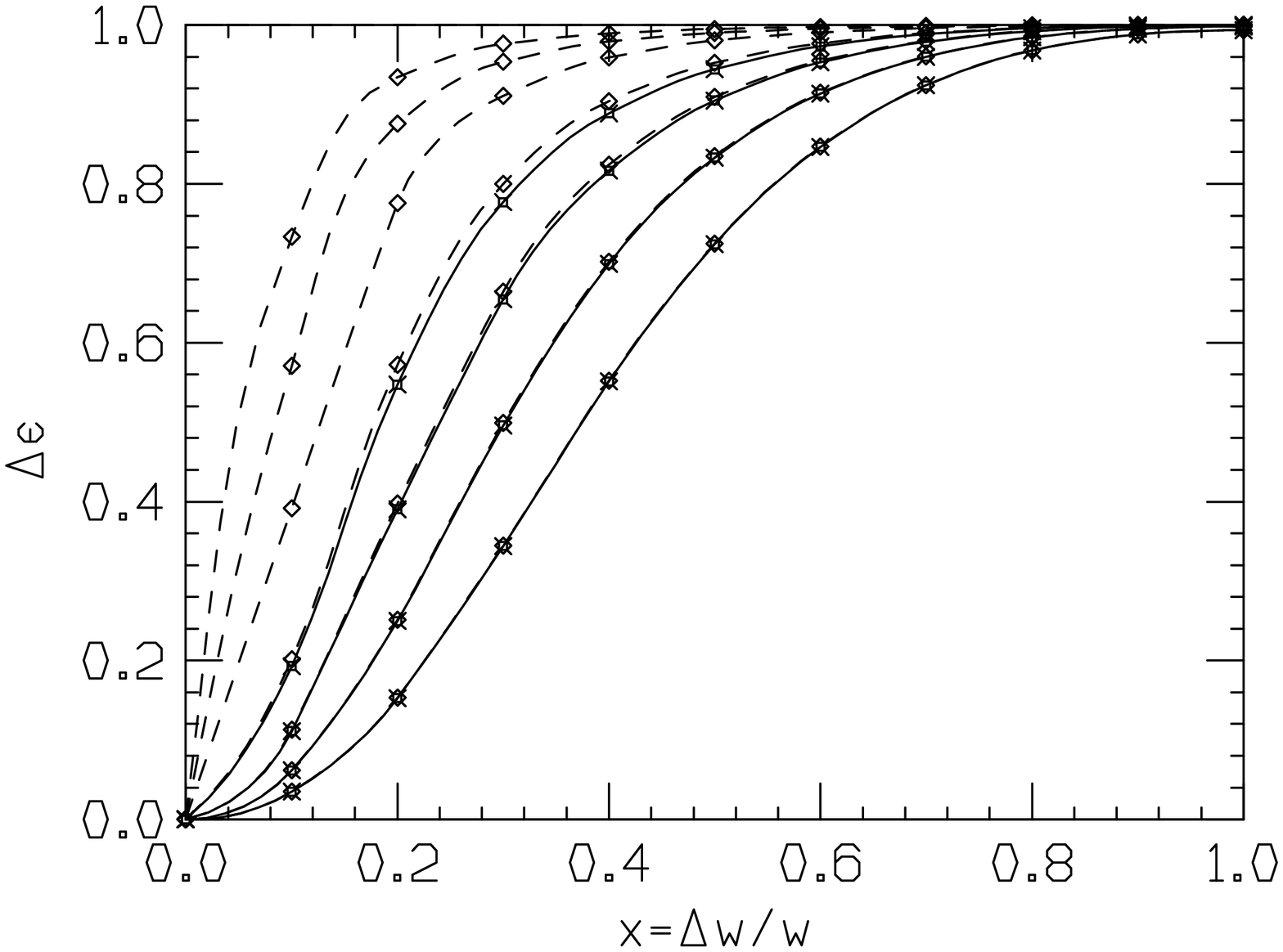}}
 \caption  { Fractional energy increase of an electron in the ring (Eq. (17)) due to rotational zero-point motion  
  as function of the fraction of the ring width $w$ occupied
 by the barrier, $x=\Delta w /w$.   The ring has internal and external radius $r_1=2$, $r_2=2.5$.
 The full-line curves are obtained for discretization in the $r$ direction with $N_r=10$, and discretization in the
 $\varphi $ direction with $N_\varphi=10, 20, 40, 80$. As $N_\varphi$ increases the curves become flatter for large $x$.
 The crosses give the actual data for each of the $10$ positions of the barrier, and are connected by smooth lines.
 The dashed lines and diamond points are obtained for a 1-dimensional chain ($N_r=1$) where the hopping between 
 the $N_\varphi$ site and site $1$ is given by Eq. (19), for 
 $N_\varphi=10, 20, 40, 80, 200, 400, 800$. Note the excellent agreement between the results for
 $N_r=10$ and $N_r=1$ for the same $N_\varphi$.}
 \label{figure2}
 \end{figure}
  
  Consider for definiteness  a ring of inner and outer radii $r_1=2$, $r_2=2.5$. Fig. 2 shows the behavior of the 
  fractional energy increase
  \beq
  \Delta \epsilon \equiv \frac{E-E_{out}}{E_{in}-E_{out}}
  \eeq
  for $N_r=10$ and various values of $N_\varphi$
  as a function of the barrier position. The horizontal axis variable $x=\Delta w/w$ takes $10$ discrete values in
  this case, with $0$ ($1$) denoting the barrier fully out (in) respectively. It can be seen that as $N_\varphi$ increases the curves 
  become increasingly flatter for large $x$. This indicates that the electron is happy in a state very similar 
  to Eq. (15) (equivalent to Eq. (1)) and is not
  exerting any `quantum pressure' to push the barrier out. We conjecture that in the continuum limit
  ($N_\varphi\rightarrow\infty$) the energy will make a crossover from Eq. (16a) to Eq. (16b) in a tiny interval
  given approximately by the variational estimate Eq. (11), which for the parameters used here
  ($r_1=2$, $r_2=2.5$) corresponds to $0\le \Delta w/w \le 1/1600$.
  
  For the ground state wavefunction, we find that to a very good approximation it factorizes as given by Eq. (14)
  with $f(i_r)$ given by Eq. (15a) and $g(i_r)$ interpolating between Eq. (15b) and a constant depending on the
  position of the barrier.  Fig. 3 shows
  results for
  \beq
  g \equiv \sqrt{\frac{N_\varphi+1}{2}}\Psi(i_r,i_\varphi)/f(i_r)
  \eeq
  as function of $\varphi=2\pi i_\varphi/N_\varphi$  with $f(i_r)$ given by Eq. (15a) and four values of the barrier position:
  $x=0$, $x=0.2$, $x=0.4$ and $x=1$.
  The $i_r$ dependence of the results is nearly indistinguishable in Fig. 3, it can only barely be discerned near the
  extremes $i_\varphi=0$ and $i_\varphi=N_\varphi$ where the different curves very slightly `fan out'.
  It can be seen that as $N_\varphi$ increases the ground state wavefunction for any value of
  $\Delta w/w$ except zero approaches the wavefunction for $x=\Delta w/w=1$ given by Eq. (15b).
  
  In addition the numerical results show that the wavefunction Eq. (18) as well as the ground state energy are
  essentially independent of the
  degree of lattice discretization in the $r$ direction ($N_r$) provided the strength of the hopping across the barrier
  is properly scaled. As the extreme case we consider a 1-dimensional chain 
  ($N_r=1$) of $N_\varphi$ sites and take the
  hopping between site $N_\varphi$ and site $1$ to have magnitude 
  \beq
  t_\varphi'=t_\varphi\sum' f(i_r)^2
  \eeq
  with $f(i_r)$ given by Eq. (15a), and 
  where the sum extends over the $i_r$ points where the barrier is absent.
  The numerical results from this calculation are shown in Fig. 3 as the points, and the results for the case
  $N_r=10$ by the lines. Similarly in Fig. 2 the numerical results from this calculation are shown as the diamonds
  and the dashed lines.
  It can be seen that the agreement between the results for $N_r=1$ and $N_r=10$ is nearly perfect
  for the ring considered ($r_1=2$, $r_2=2.5$). This no longer holds for rings where the width becomes comparable or
  larger than the inner radius. 
  For the single chain with one hopping $t_\varphi'<t_\varphi$ it can be shown analytically that 
  in the continuum limit  (i.e. $N_\varphi \rightarrow \infty$, $a_\varphi \rightarrow 0$)  the wavefunction and the energy converge to the results for the open chain ($t_\varphi '=0$),
  no matter how small the difference between $t_\varphi'$ and $t_\varphi$ is.
  
       \begin{figure}
  \resizebox{8.5cm}{!}{\includegraphics[width=8cm]{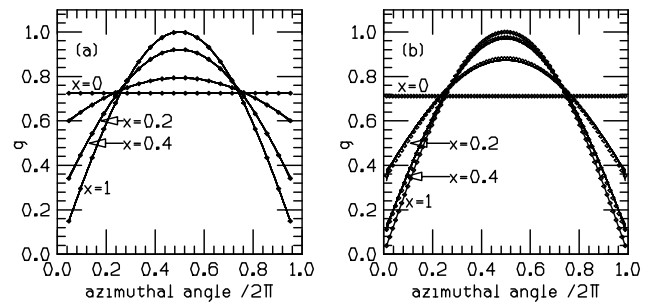}}
 \caption  { Scaled ground state wavefunction Eq. (18) versus azimuthal angle for different positions of the barrier. $x=\Delta w/w$ denotes the fraction of the ring width obstructed by the barrier. The solid lines correspond to discretization in the $r$ direction with $N_r=10$, and the points to
 $N_r=1$ with the hopping amplitude   between the $N_\varphi$ site and site $1$  given by Eq. (19). Note the excellent agreement
 between the results for $N_r=10$ and $N_r=1$. (a) are results for discretization in the azimuthal direction with
 $N_\varphi=20$, and (b) with $N_\varphi=80$. 
 Note that as $N_\varphi$ increases the wavefunction for any $x$
 approaches the wavefunction for $x=1$ (complete obstruction).  In Fig. 1(b) the results for 
 $40\%$ obstruction are nearly indistinguishable from those with $100\%$ obstruction.  }
 \label{figure2}
 \end{figure}

  In summary, these results indicate that the behavior of an electron in a ring predicted by conventional quantum 
  mechanics is very peculiar. As the barrier is gradually pulled out the electron is trying very hard to remain
  in the state where the barrier was fully in, with phase change $\pi$ around the ring, rather than 
  helping to push the barrier out
  and eliminating the phase change,
  except near the very end of the process where the barrier is almost out.

 I argue that these predictions of conventional quantum mechanics are so highly counterintuitive that perhaps they do not describe physical reality.
 When the barrier is completely in, the zero-point energy  Eq. (3) is non-zero because the angle $\varphi$ is confined to the finite
 region Eq. (4). 
 As the barrier is   pulled out, the azimuthal angle $\varphi$  remains confined to the same finite region. I  propose  that as a consequence ``zero point motion'' still 
has to exist
and the azimuthal energy Eq. (3) will be  $unchanged$. The wavefunction is no longer constrained to be the particular linear
combination Eq. (1) in the region where the barrier is absent, thus it can be either
\bmath
\beq
\psi_1(\varphi)=e^{i\varphi/2}
\eeq
or
\beq
\psi_2(\varphi)=e^{-i\varphi/2} .
\eeq
\emath
Putting the barrier back will change the wavefunction to the appropriate linear combination of (20a) and (20b) but shouldn't cost any energy because the electron in this geometry cannot ``push'' radially outward. 
The wavefunctions Eq. (20) have angular momentum in the $z$ direction
\beq
L_z=\pm \hbar /2
\eeq
and I propose that this is the minimum value of orbital angular momentum for an electron in the ring 
without the barrier when phase coherence exists,
rather than the prediction of conventional quantum mechanics $L_z=0$ (Eq. (5)).
 
If an electron is in a $linear$  box of length $L$ it  oscillates back and forth with speed $v=\hbar \pi/(m_e L)$ in its ground state. If one of the walls is not infinitely high,
or is suddenly made more transparent, or is removed very quickly, the electron will tunnel out or fly out
with the same speed $v=\hbar \pi/(m_e L)$ that it had when the wall was in place.
Similarly the electron in the case of Fig. 1(a) is orbiting back and forth in the ring with speed
$v=\hbar/2m_e r$, with $r$ the radius of the ring, and one might reasonably expect that the electron will keep this speed if the barrier is
suddenly removed or made transparent. This is consistent with the electron having orbital angular momentum of $\hbar/2$ or
$-\hbar/2$ when the barrier is no longer there, as described by the wavefunctions Eq. (20).  Or, the electron could stay in the state Eq. (1) if the
`sudden approximation' is valid and $\Psi(t=0^+)=\Psi(t=0^-)$ . Note that within conventional quantum mechanics the sudden approximation has to
break down for this situation because it is impossible to express the ground state of the electron in Fig. 1(a) as a linear combination of states of the electron
in Fig. 1(b).

Besides the considerations above, this proposal is motivated by  a prediction of the theory of hole superconductivity, proposed to describe all superconductors. It was found within that theory\cite{sm} that 
in the superconducting state electrons move in mesoscopic orbits of radius $2\lambda_L$ ($\lambda_L=$ London penetration depth) with speed
$v_\sigma^0=\hbar/(4 m_e\lambda_L)$, thus carrying orbital angular momentum $\hbar/2$. Electrons in superconductors have macroscopic phase coherence.
Thus it is natural to infer  that the angular momentum $\hbar/2$ and associated phase change of $\pi$ when the electron traverses a closed $2\lambda_L$ orbit is
an intrinsic property of the phase-coherent electron rather than a particular property of  superconductors.

More generally I point out that a non-zero ground state angular momentum provides a physical argument
 for ``quantum pressure'' and the stability of matter\cite{lieb} 
 that is absent in conventional quantum theory.
A particle of mass $m$  rotating in a circle of radius $r$ with  angular momentum $L$ has kinetic energy 
\beq
E_{kin}=\frac{L^2}{2mr^2}
\eeq
and reducing $r$ for fixed $L$   increases its kinetic energy, mimicking  the kinetic energy term in Schr\"odinger's equation
$-(\hbar^2/2m)\nabla^2$ for $L\sim\hbar$. This argument underlies the stability of Bohr's orbits, where the angular momentum
$L=n\hbar$ is always nonzero ($n\geq 1$). De Broglie was guided to his relation $p=h/\lambda$ by the argument that a finite  number of wavelengths, $n$,
should fit into the $n$-th Bohr orbit ($2\pi r=n\lambda => L=pr=n\hbar$). This physics is lost in the Schr\"odinger  equation, that allows for zero angular momentum
solutions (e.g. the $l=0$ states of hydrogen)
 subject to `quantum pressure' of unknown origin. Thus I argue  that the old Bohr-Wilson-Sommerfeld quantization rule
\beq
\oint \vec{p}\cdot d\vec{q}=nh
\eeq
with $\vec{p}$ and $\vec{q}$ canonically conjugate and $n\geq 1$ has deeper physical content than the Schr\"odinger equation that it inspired.

The wavefunctions Eq. (20) satisfy Schr\"odinger's equation with energy Eq. (3),
are continuous and have continuous derivative as the  Schr\"odinger equation requires in the absence of infinite potentials, but are not considered to be a valid description of physical reality because they are not
single-valued\cite{yang}. When the electron goes around the ring once, it ends up in a state of opposite sign, and two rounds are needed to get back to the original state.
But if the observable object is the $square$ of the wavefunction, $P(\varphi)=|\psi (\varphi)|^2$ giving the probability of finding the electron at azimuthal angle $\varphi$, all we should
require is that $P(\varphi+2\pi)=P(\varphi)$, which the wavefunctions Eq. (20) do satisfy.
Ascribing two possible values to the electron wavefunction at the same point in space may contradict our classical physical intuition, but 
not more so than the notion
 that the same  electron somehow goes through two Young slits simultaneously in creating an interference pattern, that we have come to embrace.

And of course electrons have spin $1/2$. 
At the dawn of quantum mechanics, it was attempted to represent the wavefunction of a spinning electron by   an azimuthal angle of self-rotation $\varphi$ just like 
the form Eq. (20), but this was discarded in favor of the Pauli matrix formulation precisely because of the difficulty in understanding the 
resulting double-valuedness problem\cite{pauli}.
Nevertheless the `problem' crept back in: under a spatial rotation around an axis $\hat{n}$ through angle $\theta$  a two-component spinor   $\chi$
transforms  to\cite{merzbacher}
\bmath
\beq
\chi'=U_R\chi
\eeq
\beq
U_R=e^{-\frac{i}{2}\theta \hat{n}\cdot\vec{\sigma}}
\eeq
\emath
and $U_R(\theta=2\pi)=-1$. There has not been a clear physical interpretation of this well-known result. I argue that it implies,  if we consider the symmetry transformation $U_R$ from an
`active' rather than a `passive' point of view\cite{symmetry},  that transporting  a spin $1/2$ electron in a $360^o$ circle will change the sign of its 
wavefunction, just as described by Eq. (20).

The $\pi-$phase shift that a spinor acquires under a $2\pi$ rotation Eq. (24)  is not just a mathematical artifact, rather it is expected to be real and observable\cite{susskind}.
It can be interpreted as a geometric phase, just like  the Aharonov-Bohm (AB) phase\cite{ab}, as discussed by Berry\cite{berry}.
Experiments have shown that electrons orbiting around a magnetic flux tube do acquire the AB phase\cite{abexp}, similarly I  argue electrons will acquire their
`geometric phase' of $\pi$ proposed here each time they orbit around the ring, as described by the wavefunctions Eq. (20).

   \begin{figure}
 \resizebox{8.5cm}{!}{\includegraphics[width=8cm]{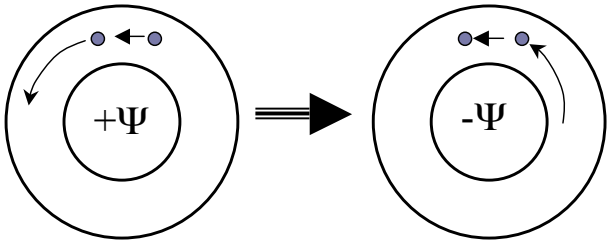}}
 \caption  { Interchanging the position of two identical fermions is topologically equivalent to rotating one around the other\cite{spinstat2}.
 The two-electron wavefunction changes its sign, both according to conventional quantum mechanics and to the
 physics proposed here.
}
 \label{figure2}
 \end{figure}

Furthermore, the topological equivalence between interchanging two spin-$1/2$  fermions and rotating one around the other 
depicted  in Fig. 4 has been 
noted and pointed out repeatedly\cite{spinstat1,spinstat2,spinstat3,wilc}. Both processes give rise to a 
$(-1)$ factor in the two-electron wavefunction, the former one is interpreted as originating in the   antisymmetry of fermion wave functions and the latter one as arising  due to the
spinor transformation Eq. (24). This gives a rationale to  the well-known spin-statistics relationship\cite{spinstat2}. 
Here I  propose that the (-) sign arising when one electron  loops  around another in a ring, thus interchanging their positions,  is  due to the single fermion
going around the circle, whether or not the other fermion is present.

Which of the two wavefunctions Eq. (20) will the electron choose? In the absence of applied magnetic field it is natural to expect that the rotational zero point motion will be
the same as that predicted for superconductors\cite{sm}, namely $\vec{v}_\sigma \parallel ( \hat{r}\times \vec{\sigma})$ for the velocity direction of the electron of spin $\vec{\sigma}$ perpendicular
to the plane of the ring and $\hat{r}$ pointing radially outward. In other words, an electron with spin pointing into (out of) the paper in Fig. 1(b) will rotate
counterclockwise (clockwise). This is the lowest energy state dictated by spin-orbit coupling in the presence of an outward-pointing electric field. 
Quite generally, because electrons are lighter than protons, they tend to move outward from compensating positive charge and thus experience 
outward-pointing electric fields.

The conventional Bohr-Wilson-Sommerfeld quantization rule Eq. (23) needs to be modified to reflect this physics, to read
\beq
\oint \vec{p}\cdot d\vec{q} =(n\pm \frac{1}{2})h
\eeq
with the $+$ ($-$) sign corresponding to spin orientation  opposite to (the same as) the direction given by the right-hand rule in traversing the integration circuit,
and $n$ an integer. It is interesting to note that Eq. (25) (with the $(+)$ sign and $n\geq 0$)  gives the correct energy for the harmonic oscillator including its
zero point motion, which the conventional rule Eq. (23) does not. 
With respect to Bohr's semiclassical model of hydrogen, note that it does not take into account spin. Taking the average of Eq. (25) over
both spin orientations will give back the conventional rule Eq. (23) and yield the correct  answers  for the energy levels of hydrogen. 

Imagine standing at a point in the ring watching an electron fly by. How can you tell when the electron returns after encircling the loop whether or
not it has retained phase coherence? In the scenario proposed here, the electron reappearing with wavefunction of opposite sign is
the telltale signature of phase coherence.   Thus the ``closing'' of the wavefunction 
emphasized by A.V. Nikulov as  characterizing  quantum coherence\cite{nikulov} can be detected $locally$ as opposed to the conventional scenario
that requires knowledge of the entire wavefunction of the electron around the loop\cite{nikulov}.

What differentiates a phase-coherent superconductor from a phase-coherent normal mesoscopic ring? In both cases the individual electron wave function changes sign in
going around the loop. But in the superconductor electrons are paired, and $(-1)\times(-1)=1$, so the Cooper pair wave function does not change sign.
This allows for the establishment of macroscopic phase coherence in the superconductor by locking the phases of different Cooper pairs, and it is not possible
for unpaired electrons because of random occurrence of positive and negative signs.

Consider now the states of several non-interacting electrons of the same spin in a ring of length $L$. The states Eq. (20) correspond to the electron having wavevector
$k=\pm \pi/L$, as shown in Fig. 5(a). The electron with spin in the $\hat{z} \equiv \hat{r}\times\hat{\varphi}$ direction will occupy the $k=-\pi/L$ state due to the
spin-orbit interaction, orbiting with angular momentum pointing in the $-\hat{z}$ direction. For two electrons of the same spin, the $k$ values shift to become
the same as conventionally, as shown in Fig. 5(b), i.e. $k=0$, $k=\pm 2\pi /L$. This is easily seen from the equivalence between interchanging fermions and
looping one around the other with a $\pi$ phase shift depicted in Fig. 4. Thus the allowed $k$-states shift back and forth as the number of electrons switches
between odd and even as shown in Fig. 5. Our theory differs from the conventional theory in that the ground state is always nearly degenerate
(degenerate if the spin-orbit interaction is ignored) rather than alternating between degenerate and non-degenerate for odd and even number of
electrons. Note that all the states depicted in Fig. 5 carry a charge current, while in the conventional theory only half 
the states (those with even number of electrons) would.

To add the opposite spin electrons to Fig. 5 we simply place them at the $k$-points that are the mirror image across the vertical direction of the states
occupied by the up electrons. Thus a system with equal number of up and down electrons will carry a spin current but no charge current, with the
net orbital angular momentum of electrons of each spin in opposite direction to their spin angular momentum.
 \begin{figure}
 \resizebox{8.5cm}{!}{\includegraphics[width=8cm]{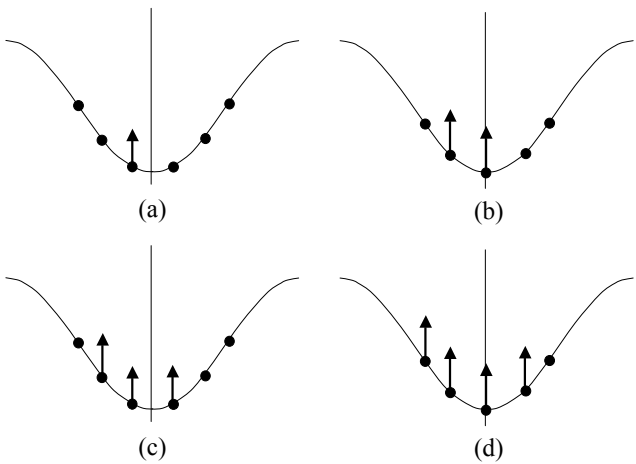}}
 \caption  {States for non-interacting identical fermions in a ring of length $L$. The dots are spaced at distance $\Delta k=2\pi/L$.
 For an even number of fermions there is a state at $k=0$, for an odd number of fermions the lowest $k$ is $\pm \pi/L$.
 The degeneracy in the ground states is broken by the spin-orbit interaction.}
 \label{figure2}
 \end{figure}
 In the presence of magnetic flux $\Phi$, the allowed wavevectors for the electrons in a ring of length $L$ are, for the case of even and odd number of 
electrons of a given spin respectively
\bmath
\beq
k_n=\frac{2\pi}{L} (n+\frac{\Phi}{\Phi_0})
\eeq
\beq
k_n=\frac{2\pi}{L} (n\pm \frac{1}{2} +\frac{\Phi}{\Phi_0})
\eeq
\emath
with $\Phi_0=hc/e$ the flux quantum.

 The ``tireless electrons''\cite{imry} of Fig. 5 will have a   stronger tendency to move  around and create persistent currents than conventional
 electrons, where half of the states do  not carry current. 
 There have been many experiments on persistent currents in mesoscopic rings\cite{imry, exp1,exp2,exp3} over the years and the field has been mired in controversy.
 Although it is not apparent in many of the papers written on the subject, the underlying theme is the surprisingly large persistent currents
 that are typically observed, larger by up to two orders of magnitude from what is theoretically expected.
 Within the physics proposed here, mesoscopic rings will carry a {\it spin current} in the absence of applied magnetic field, and an applied magnetic flux will simply
 slow down one of the components of the spin current and speed up the other thus giving rise to a charge current, rather than creating  the
 charge  current from scratch.  It should be in principle possible (though experimentally challenging) to detect the spin current in the absence of applied magnetic 
 field through the small electric field that it creates. It should be   easier in superconducting rings as discussed in \cite{sm}.
 The physics discussed here may also be related to experimental signatures of spin currents
 in surface states observed in a variety of
 materials in recent years\cite{gold,ti}.
 The quantum number shift in the Bohr-Wilson-Sommerfeld quantization rule proposed here (Eq. (25)) may   provide an
 explanation for the puzzling experimental observations of Nikulov and coworkers on asymmetric 
 superconducting rings\cite{nikulov2}.

There is in fact no reason to restrict the physical arguments presented here to the ring topology.  
They suggest that quite generally  the origin of electronic  `quantum pressure'  in nature, manifest in the fact that electrons tend to   expand their wavefunction radially 
as far as possible, is that   they  undergo 
{\it zero-point rotational motion} in the  region of space that they have available, with angular momentum
$\hbar/2$, just  as predicted for electrons    in superconductors\cite{efm,sm}.    This spinning zero-point orbital motion 
originates in the two-valuedness of the electron wavefunction and   carries the same magnitude of angular momentum as, and opposite direction to,
the intrinsic electron spin,   which itself can be represented  by a mass $m_e$ orbiting at speed $c$ in
a circle of  radius
$r_q=\hbar/(2m_e c)$. 
I do not explain here the   origin of this zero-point rotational agitation, which presumably derives from 
the topological structure of space-time itself.  
Schr\"odinger's equation, while undoubtedly correct for a large number of  physical situations, does not describe this physics,
nor does Dirac's equation in its current form.
If correct it is evident  that this physics has profound implications for the understanding of matter. In particular I discuss
elsewhere\cite{aromatic} that it leads to the expected existence of ground state spin currents in aromatic ring molecules, ubiquitous in biological matter.
This physics   will  generally lead to   less `inert' structures than the conventional understanding, and 
 may ultimately explain  questions as general as  the `elan vital' and how the universe avoids heat death.

It is likely that  the concepts discussed in this paper have  connections and overlaps with a variety of concepts that have been discussed in the condensed matter and particle physics  literature in recent years
such as Dirac monopoles,   anyons, Chern numbers, flux phases, composite fermions, TKNN invariant, 
quantum spin Hall effect, topological superfluids, topological insulators, fiber bundles, 
Berry phase, Aharonov-Casher phase, vortices, strings, dyons, skyrmions, etc. I have not elucidated these connections  in detail
 and apologize for not citing possibly relevant references, but believe that
the concrete physics proposed in this paper has not been proposed before. Connections with other related work should yield interesting insights and   further progress
in understanding. 

\noindent {\bf Note added:} After completion of this paper it came to  my attention that the possibility of a double-valued wavefunction for the electron has been considered by
various workers in the past\cite{schr,pauli2,merz}. However the possibility that this may explain the origin of   `quantum pressure' has not been suggested before to my knowledge.

\acknowledgements
I have benefitted from stimulating correspondence with A.V. Nikulov and from reading his
insightful papers, and from stimulating criticism by D.C. Mattis that led to improvements on the
first version of the manuscript.


\begin{references}
     \bibitem{parr} L.C. Snyder and R.G. Parr, J. Chem. Phys. {\bf 34}, 1661 (1961).
  \bibitem{sm}  J.E. Hirsch, Europhys. Lett. {\bf 81}, 67003 (2008) and references therein.
     \bibitem{lieb} E.H. Lieb, Rev. Mod. Phys. {\bf 48}, 553 (1976), Sect. 1.
               \bibitem{yang} C.N. Yang, in Proc. Int. Symp. Foundations of Quantum Mechanics, Tokyo, 1983, p. 5 states: ``We emphasize that
               to challenge the single valuedness requirement of the wavefunction is to challenge the very foundation of quantum mechanics itself''.
               \bibitem{pauli} W. Pauli, Z. f\"ur Physik A {\bf 43}, 601(1927).
                   \bibitem{merzbacher} E. Merzbacher, `Quantum Mechanics', Third Edition, John Wiley and Sons, Inc, New York 1998, Chpt. 12.
                    \bibitem{symmetry} R. Hagedorn, 	Il Nuovo Cimento {\bf 12}, Suppl. 1, p. 73 (1959).
                                    \bibitem{susskind} Y. Aharonov and L. Susskind, Phys. Rev. {\bf 158}, 1237 (1967).
                      \bibitem{ab} Y. Aharonov and D. Bohm, Phys. Rev. {\bf 115}, 485 (1959).
                      \bibitem{berry} M.V. Berry, Proc. R. Soc. London A{\bf 392}, 45 (1984).
                      \bibitem{abexp} N. Osakabe et al, Phys. Rev. A {\bf 34}, 815 (1986).
                    \bibitem{spinstat1} M.V. Berry and J.M. Robbins, Proc. R. Soc. London A {\bf 453}, 1771 (1997).
                     \bibitem{spinstat2}R.P. Feynman, ``The reasons for antiparticles'', in 
                     ``Elementary Particles and the Laws of Physics. The 1986 Dirac Memorial Lectures'', ed. by R.P. Feynman and S. Weinberg,
                     Cambridge University Press, New York 1987, p. 56-59.
                      \bibitem{spinstat3} D.E. Neuenschwander, Am. J. Phys. {\bf 62}, 972 (1994).
                      \bibitem{wilc} F. Wilczek, Phys. Rev. Lett. {\bf 48}, 1144 (1982).
                      \bibitem{nikulov} A.V. Nikulov,  arXiv:0812.4118v1 (2008); Phys. Rev. B{\bf 64}, 012505 (2001).
                      \bibitem{imry} J. Imry, Physics {\bf 2}, 24 (2009) and references therein.
                      \bibitem{exp1} H. Bluhm et al, Phys. Rev. Lett. {\bf 102}, 136802 (2009) and references therein.
                        \bibitem{exp2} A. C. Bleszynski-Jayich et al, Science  {\bf 326}, 272 (2009) and references therein.
                        \bibitem{exp3} U. Eckern and P. Schwab, J. Low Temp. Phys. {\bf 126}, 1291 (2002) and references therein.
                                    \bibitem{gold} S. LaShell, B. A. McDougall, and E. Jensen, Phys. Rev. Lett. {\bf 77},  3419  (1996).
                        \bibitem{ti} D. Hsieh et al, Science {\bf 323}, 919 (2009).
                        \bibitem{nikulov2} V.L. Gurtovoi et al, JETP {\bf 105}, 262 (2007).
                        \bibitem{efm}      J.E. Hirsch, Jour. of Superconductivity and Novel Magnetism {\bf 23}, 309 (2010).
                        \bibitem{aromatic} J.E. Hirsch, Mod. Phys. Lett. {\bf B4}, 739 (1990) and 
                        doi:10.1016/j.physleta.2010.07.023 (2010).
                        \bibitem{schr} E. Schr\"odinger, Ann. der Physik {\bf 32}, 49 (1938).
                        \bibitem{pauli2} W. Pauli, Helv. Phys. Acta {\bf 12}, 147 (1939).
                        \bibitem{merz} E. Merzbacher, Am. J. Phys. {\bf 30}, 237 (1962).
            
      
  \end{references}
\end{document}